\documentclass[10pt]{iopart}
\pdfoutput=1
\usepackage{graphicx}
\usepackage{amssymb}
\usepackage{xcolor}
\usepackage{iopams} 

\usepackage[colorlinks,linkcolor=phthaloblue,urlcolor=blue,citecolor=phthaloblue]{hyperref}
\usepackage{color}
\definecolor{phthaloblue}{rgb}{0.0, 0.06, 0.54}

\usepackage{bm}

\eqnobysec

\newcommand{\beq}{\begin{equation}}
\newcommand{\eeq}{\end{equation}}
\newcommand{\be}{\begin{equation*}}
\newcommand{\ee}{\end{equation*}}
\newcommand{\beqa}{\begin{eqnarray}}
\newcommand{\eeqa}{\end{eqnarray}}
\newcommand{\bea}{\begin{eqnarray*}}
\newcommand{\eea}{\end{eqnarray*}}

\def\stackunder#1#2{\mathrel{\mathop{#1}\limits_{#2}}}
\newcommand{\bigmean}[1]{\left\langle#1\right\rangle}
\newcommand{\mean}[1]{\langle#1\rangle}
\newcommand{\bigac}[1]{\left(#1\right)}
\newcommand{\lap}[1]{\mathrel{\mathop{\cal L}\limits_{#1}^{}}}

\newcommand{\prob}{\mathbb{P}}

\newcommand{\ii}{{\rm i}}

\newcommand{\la}{{\lambda}}

\newcommand{\m}[1]{{\bm#1}}

\newcommand{\bra}[1]{\langle#1\vert}
\newcommand{\ket}[1]{\vert#1\rangle}
\newcommand{\braket}[2]{\langle#1\vert#2\rangle}
\newcommand{\braopket}[3]{\langle#1\vert#2\vert#3\rangle}

\begin{document}
\title{Occupation time of a renewal process coupled to a discrete Markov chain}

\author{Th\'{e}o Dessertaine$^1$$^,$$^2$, Claude Godr\`{e}che$^3$ and Jean-Philippe Bouchaud$^2$$^,$$^4$}
\address{$^1$LadHyX UMR CNRS 7646, \'Ecole Polytechnique, 91128 Palaiseau Cedex, France}
\address{$^2$Chair of Econophysics \& Complex Systems, \'Ecole Polytechnique, 91128 Palaiseau Cedex, France}
\address{$^3$Universit\'e Paris-Saclay, CNRS, CEA, Institut de Physique Th\'eorique,
91191~Gif-sur-Yvette, France}
\address{$^4$Capital Fund Management, 23 rue de l'Universit\'{e}, 75007 Paris, France\medskip}

\begin{abstract}
A semi-Markov process is one that changes states in accordance with a Markov chain but takes a random amount of time between changes. 
We consider the generalisation to semi-Markov processes of the classical Lamperti law for the occupation time of a two-state Markov process. 
We provide an explicit expression in Laplace space for the distribution of an arbitrary linear combination of the occupation times in the various states of the process. 
We discuss several consequences of this result.
In particular, we infer the limiting distribution of this quantity rescaled by time in the long-time scaling regime, as well as the finite-time corrections to its moments.
\end{abstract}

\eads{\mailto{theo.dessertaine@polytechnique.edu}, \mailto{claude.godreche@ipht.fr}, \mailto{jean-philippe.bouchaud@academie-sciences.fr}}

\maketitle

\section{Introduction}
\label{sec:intro}

Studies of the occupation time of stochastic processes have a long history, starting with the investigation by L\'evy of the fraction of time spent by Brownian motion above zero or of the fraction of time where the first player is ahead of the second, in repeated tossings of a coin \cite{levy}.
The limiting density of this fraction of time is the U-shaped arcsine law, with a minimum at $1/2$ and infinite tails at $0$ and $1$ (see \cite{feller49,pitman} for a summary of \cite{levy}).
These founding investigations were followed by many subsequent studies
\cite{feller49,kac,darling,lamperti,takacs,kesten62} and the topic is now a classic in probability theory (see \cite{pitman2,james} for reviews).

Only lately was this topic revisited in the physics community, motivated, in particular, by studies on phase persistence for self-similar coarsening systems, such as breath figures \cite{marcos}, Ising spin systems quenched from high temperature to zero temperature (or more generally to a temperature below the critical temperature) \cite{dbg94,bdg94}, the diffusion equation evolving from a random initial condition \cite{majdiffu,derrdiffu}, to name but a few.
The statistics of the occupation time for Ising spins systems, the voter model and diffusive persistence were addressed in \cite{dorn}, while \cite{newman} is entirely devoted to this last subject. 
However, obtaining a complete solution to the question of the statistics of the occupation time for these extended systems is however currently out of reach.
Partial analytical studies, as well as
numerical or approximate theoretical treatments, allow nevertheless to make progress \cite{dorn,newman,drouffe,balda,newman2,drouffe2}. 
A noticeable series of works on the occupation time of the voter model \cite{cox1,cox2,cox3,brams,cox4} was a source of inspiration for the endeavour made in \cite{dorn} on this topic.
A study on the same issues was further continued in \cite{maillard}.

The above mentioned works \cite{dorn,newman} were followed by investigations on the statistics of the occupation time for simpler systems, more amenable to exact analysis, and closer to the main stream of probabilistic studies \cite{balda,bauer,dhar,smedt,gl2001,moving,bray,barkai4}.
We refer the reader to \cite{revue} for subsequent references and to \cite{angel,bridge,bress1,burkh,barkai18,radice} for more recent works.

The study presented below belongs to the same vein and gives a generalisation to multistate Markov processes of the Lamperti law for the occupation time of a two-state Markov process \cite{lamperti}, a problem also considered in \cite{barkai1, barkai2, barlow}. Our motivation comes from a recent work by two of us \cite{theo} on cone-wise linear dynamics in large dimensions. 
Each ``cone'' is characterised by a stability matrix chosen to be from the Gaussian Orthogonal Ensemble. 
The selected cone is determined by the direction of the dynamically evolving vector with respect to a fixed set of vectors. Because of the random nature of the stability matrix, the cone-switching process can be described, for large dimensions, as a semi-Markov process, with a power-law distribution of switching times \cite{theo}. 

\begin{figure}[!ht]
\begin{center}
\includegraphics[angle=0,width=.9\linewidth,clip=true]{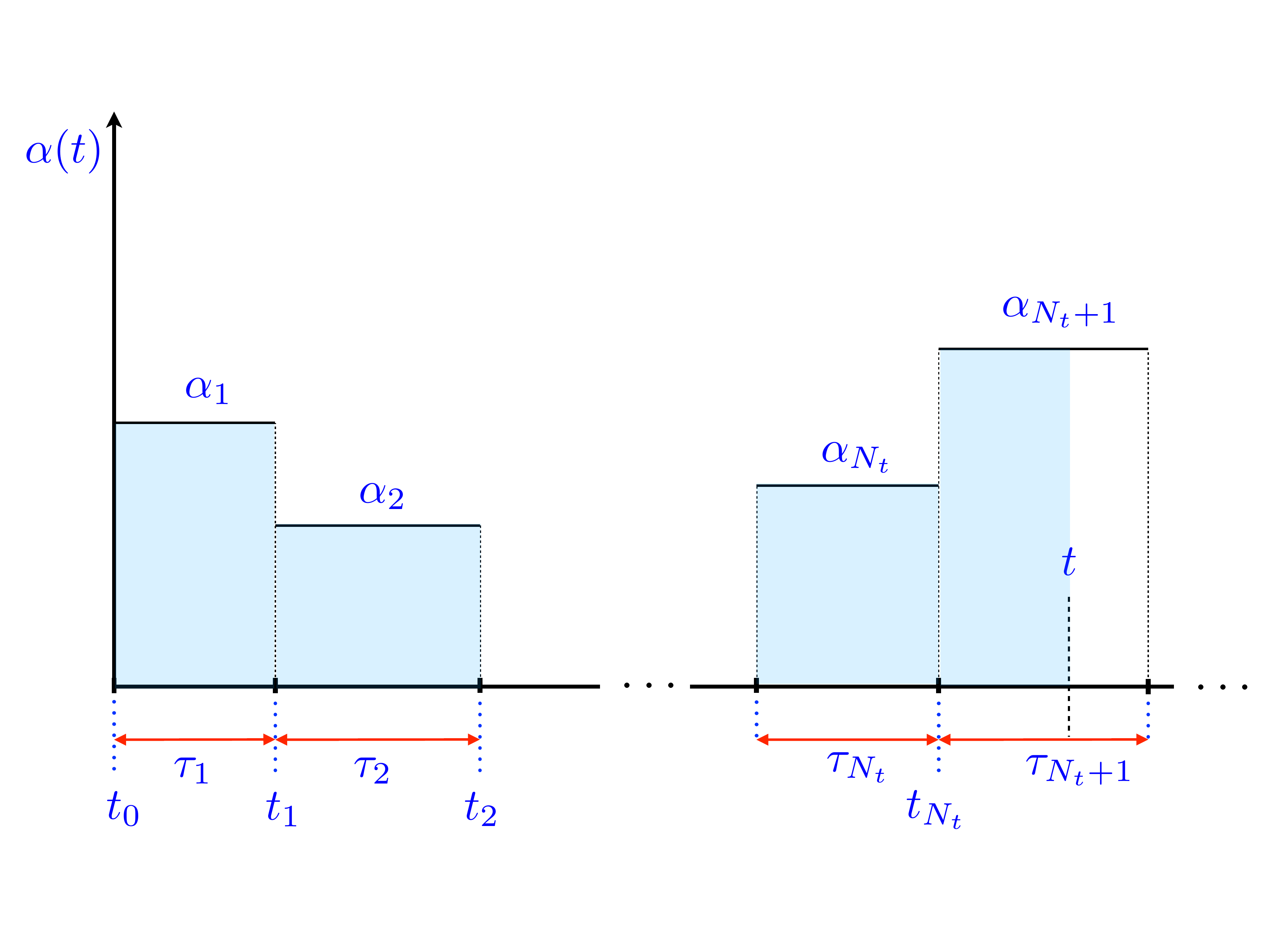}
\caption{\small
The process $\alpha(t)$ is constant in each time interval $\tau_1,\tau_2,\dots$, and takes one of the $q$ values $a_1,\dots,a_q$.
The sum $S_t$, defined in (\ref{eq:defSt}), is the linear combination (\ref{eq:Tt}) of occupation times up to time $t$ of the process $\alpha(t)$ in these various states.
It is depicted by the area in blue.
The last time interval, $t-t_{N_t}$, entering the definition of $S_t$, is the age of the current interval at time $t$.}
\label{fig:occ}
\end{center}
\end{figure}

We consider an irreducible Markov chain $\alpha_1,\alpha_2,\dots$, with discrete state space $\{a_j,j=1,2,\dots,q\}$,
and transition matrix 
\beq\label{eq:markov}
P_{ij}=\prob(\alpha_{n+1}=a_j|\alpha_n=a_{i}).
\eeq
In addition, we consider a sequence of time intervals $\tau_1,\tau_2,\dots$, which 
represent the time spent in the states $\alpha_1,\alpha_2,\dots$
More specifically, the jumps occur at the random epochs of
time $t_{1},t_{2},\ldots$, from some time origin $t_0=0$ (see figure \ref{fig:occ}). 
The intervals of time between jumps, 
$\tau_{1}=t_{1},\tau_{2}=t_{2}-t_{1},\ldots $, are independent and identically
distributed random variables with a common density $\rho (\tau)$, thus forming a renewal process
\cite{cox,coxmiller,feller2,grimmett}. 
The process defined by the pairs $(\alpha_n,t_n), n=1,2,\dots$,
is known as a Markov renewal process in the mathematical literature,
while the process defined as
\beq\label{eq:semi}
\alpha(t)=\alpha_n,\qquad t_{n}<t<t_{n+1}
\eeq
is a simple example of a semi-Markov process \cite{levy54,smith55,kesten62,cinlar,ross}. 
The latter is not Markovian except at the epochs of jumps.
As set forth in \cite{ross}, a semi-Markov process is one that changes states in accordance with a Markov chain but takes a random amount of time between changes.
If the latter is exponentially distributed, the process becomes an ordinary Markov chain in continuous time.

The purpose of this paper is to investigate the statistics of the sum
\beq\label{eq:defSt}
S_t=\int_{0}^t{\rm d} u\, \alpha(u)=\alpha_1\tau_1+\cdots+\alpha_{N_t}\tau_{N_t}+\alpha_{N_t+1}\Big(t-\sum_{i=1}^{N_t}\tau_i\Big),
\eeq
where $N_t$ is the random number of jumps between $0$ and $t$.
More precisely, we shall investigate the limiting distribution $f_M(x)$
of the fraction $M_t=S_t/t$ in the long-time limit, where
\be
M=\lim_{t\to\infty}M_t=\lim_{t\to\infty}\frac{S_t}{t}=\lim_{t\to\infty}\frac{1}{t}\int_{0}^t{\rm d } u\, \alpha(u),
\ee
that is, the distribution of the temporal mean of $\alpha(t)$, when the density $\rho(\tau)$ has a power-law tail (\ref{eq:power}), with index $\theta<1$.

There are several possible interpretations to the quantities $S_t$ or $M_t$.
The first one is in terms of occupation times.
To simplify, consider the case where the number of states is $q=2$ with $a_1=0$ and $a_2=1$.
Then $S_t$ is the occupation time of state $a_2$ (i.e., the time spent in this state), up to time $t$.
More generally, $S_t$ is the linear combination of the occupation times of the process in the various states $a_1,\dots,a_q$,
\beq\label{eq:Tt}
S_t=a_1T_t^{(1)}+a_2T_t^{(2)}+\cdots+a_qT_t^{(q)},
\eeq
where $T_t^{(j)}$ is the occupation time in state $a_j$ (i.e., the time spent in this state), up to time $t$,
with
\be
\sum_{j=1}^q T_t^{(j)}=t.
\ee
Equivalently, $M_t$ is the mean of $a_1,a_2,\dots,a_q$ weighted by the fractions of time $T_t^{(1)}/t,T_t^{(2)}/t,\dots,T_t^{(q)}/t$, spent in these various states.

A second interpretation is in terms of a one-dimensional random walk in continuous time.
Let $\alpha_1,\alpha_2\dots$ be the respective positions of the walker during the time intervals $\tau_1,\tau_2,\dots$, with $\alpha=1,2,\dots,q$.
Then $M_t=S_t/t$ is the mean position of this walker up to time $t$.
Alternatively, let $\alpha_1,\alpha_2\dots$ be the respective velocities of the walker during the time intervals $\tau_1,\tau_2,\dots$.
Then $S_t$ is the position at time $t$ of this walker and $M_t$ is its mean speed.

Likewise, if $\alpha$ is a Potts spin with $q$ states $a_1,\dots,a_q$, then $M_t$ represents the mean magnetisation up to time $t$. In the context of cone-wise linear systems, $M_t$ is, for large times, the Lyapunov exponent of the dynamics \cite{theo}.

To anticipate on what follows, a natural question is to know whether the process is ergodic, i.e., whether the distribution of the mean $M_t$ becomes narrow around $\mean{\alpha}$ in the long-time limit, or otherwise stated, is self-averaging. 
As we shall see, the answer depends on the nature of the distribution of waiting times $\tau_1,\tau_2,\dots$. 
Finally, note that the sum $S_t$ is a particular instance of what is known in the mathematical literature as a renewal-reward process (see e.g., \cite{grimmett,ross} for details).

In the specific case where the distribution of waiting times $\rho(\tau)$ has a power-law tail (\ref{eq:power}), with index $\theta<1$,
 we find that, 
 within each sector $a_i<x<a_{i+1}$, $(i=1,2,\dots,q)$,
\beqa\label{eq:central}
\fl
f_M(x)=
\frac{\sin{\pi\theta}}{\pi}
\frac{\big(\sum_{j\leq i}\pi_j\,\delta_j^{\theta-1}\big)\big(\sum_{j> i}\pi_j\,\delta_j^{\theta}\big)+\big(\sum_{j\leq i}\pi_j\,\delta_j^{\theta}\big)\big(\sum_{j> i}\pi_j\,\delta_j^{\theta-1}\big)}
{\big(\sum_{j\leq i}\pi_j\,\delta_j^{\theta}\big)^2+\big(\sum_{j> i}\pi_j\,\delta_j^{\theta}\big)^2+2\cos{\pi\theta}\big(\sum_{j\leq i}\pi_j\,\delta_j^{\theta}\big)\big(\sum_{j> i}\pi_j\,\delta_j^{\theta}\big)},
\nonumber\\
\eeqa
where $\delta_j=\left|x-a_j\right|$, and $\pi_j$ is the $j-$th component of the stationary measure.
For $x$ outside the range of values $(a_1,a_q)$, $f_M(x)=0$.
This result is universal with respect to $\rho(\tau)$, i.e., independent of the details of this distribution. 
The same expression was obtained in \cite{barkai1,barkai2} as a generalisation of the Boltzmann distribution for systems showing weak ergodicity breaking \cite{web} (see also \cite{barlow})%
\footnote{We shall come back to \cite{barkai1,barkai2,barlow}, which are closely related to the present work, in section \ref{sec:discussion}.}. 
For a uniform stationary probability measure over two states, the expression (\ref{eq:central}) recovers the classic Lamperti law \cite{lamperti}
(see (\ref{eq:lamperti})).
For cone-wise linear systems, (\ref{eq:central}) gives the distribution of the Lyapunov exponent when the mean waiting time within each cone is infinite. 
The dynamical system investigated in \cite{theo} is among the few known examples where the Lyapunov exponent is not self-averaging (see e.g., \cite{barkai-korabel} for a similar mechanism in the context of Pomeau-Manneville maps).

\section{Renewal processes: a brief reminder}
\label{sec:renewal}

As mentioned above, jumps (or renewals) occur at the random epochs of
time $t_0,t_{1},\ldots$. 
We take the origin of time at $t_0=0$.
Intervals of time between two jumps,
$\tau_{1}=t_{1},\tau _{2}=t_{2}-t_{1},\ldots $, are iid random variables with common density $\rho (\tau )$.
In other words, $\tau_2,\tau_3,\dots$ are independent copies of the generic waiting time $\tau_1$.
The number of jumps which occurred between $0$ and $t$, denoted by 
$N_{t}$, is the random variable for the largest $n$ for which $t_{n}\le t$, with
\be
t_n=\tau_1+\cdots+\tau_n.
\ee
With this definition, if there is no jump between 0 and $t$, then $N_t=0$.
The probability of such an event, or survival probability (or yet persistence probability),
reads: 
\be
q(t)=\prob(\tau_1 >t)=\int_{t}^{\infty }{\rm d}\tau\,\rho (\tau). 
\ee
The density $\rho(\tau)$ can be either a narrow distribution with all
moments finite, in which case the decay of $q(t)$, as $t\rightarrow \infty $,
is faster than any power law, or a distribution characterised by a
power-law tail with index $\theta>0 $ 
\beq\label{eq:pers}
q(t)=\int_{t}^{\infty }{\rm d}\tau\,\rho (\tau)
\approx \left( \frac{\tau_{0}}{t}\right) ^{\theta }, 
\eeq
hence
\beq\label{eq:power}
\rho(\tau)\approx \frac{c}{\tau^{1+\theta}},
\eeq
where $\tau_{0}$ is a microscopic time scale and $c=\theta\tau_0^\theta$ is the tail parameter.
If $\theta <1$ all moments of 
$\rho(\tau) $ are divergent, if $1<\theta <2$, the first moment 
$\mean{\tau_1}$ is finite but higher moments are divergent, and so on. 
In Laplace space, where $s$ is conjugate to $\tau$, for a narrow distribution we have 
\be
\lap{\tau}\rho (\tau)=\hat{\rho}(s)=\int_0^\infty{\rm d}\tau\, \e^{-s \tau}\rho(\tau)
\stackunder{=}{s\rightarrow 0}
1-\mean{\tau_1} s+\frac{1}{2}
\bigmean{\tau_1^2} s^{2}+\cdots .
\ee
For a broad distribution, (\ref{eq:pers}) yields 
\beq\label{eq:spetit}
\hat{\rho}(s)\stackunder{\approx}{s\rightarrow 0}\left\{ 
\begin{array}{ll}
1-A\,s^{\theta } & (\theta <1) \\ 
1-\mean{\tau_1} s+A\,s^{\theta } & (1<\theta <2),
\end{array}
\right.
\eeq
and so on, where $A=c|\Gamma (-\theta )|$.
From now on, unless otherwise stated, we shall only consider the case $0<\theta<1$. When $\theta > 1$, the process becomes ergodic, in the sense that $M= \lim_{t \to \infty} S_t/t$ converges to the ergodic mean, with possibly non trivial corrections when $t$ is large but finite (see section \ref{sec:moments}).

The last time interval involved in the sum (\ref{eq:defSt}) is the backward recurrence time $B_{t}=t-t_{N_t}$, i.e., the length of time measured backwards from $t$ to the last jump before $t$ (see figure \ref{fig:occ}), 
where $t_{N_t}$, the time of occurrence of this last jump, is the sum of a random number of random variables
\be
t_{N_t}=\tau_{1}+\cdots +\tau_{N_t}.
\ee
The backward recurrence time has also the interpretation of the age of the current, unfinished, interval at time $t$.
The statistics of the quantities $N_t,t_{N_t},B_t$ is investigated in detail in~\cite{gl2001}, which also contains relevant references on renewal processes.

\section{Distribution of the sum $S_t$ when $\alpha_1,\alpha_2,\dots$ are iid random variables}
\label{sec:iid}
We start our study with the simpler case where the random variables $\alpha_1,\alpha_2,\dots$ in (\ref{eq:defSt})
are independent and identically distributed with common distribution $f_\alpha(a)$, as a preparation for the more elaborate situation where these random variables form the Markov chain defined in (\ref{eq:markov}), and also because this case has an interest in itself.
The distribution $f_\alpha(a)$ is either a density, 
\be
f_\alpha(a)=\frac{{\rm d }}{{\rm d } a}\prob(\alpha\leq a),
\ee
for continuous random variables, or is given by
\beq\label{eq:faaD}
f_\alpha(a)=\sum_{j=1}^qp_j\delta(a-a_j),\qquad \sum_{j=1}^qp_j=1,
\eeq
in the discrete case.
As we shall see, in the long-time scaling regime, the distribution of the fraction $M=\lim_{t\to\infty}S_t/t$ for this latter case is the same as for the Markov renewal process investigated in section \ref{sec:markov}.

\subsection{The distribution of $S_t$}

The methods used in \cite{gl2001} for the computation of the distribution of the occupation time of a two-state process can be easily extended to the case of the multistate process at hand.

The joint probability density of $S_t$ and $N_t$ reads
\be
f_{S_t,N_t}(t,y,n)=\frac{{\rm d }}{{\rm d } y}\prob(S_t\leq y,N_t=n),
\ee
from which the density of $S_t$ is obtained by summing upon $n$
\be
f_{S_t}(t,y)=\sum_{n\ge0}f_{S_t,N_t}(t,y,n).
\ee
The computation of this density can be made in Laplace space.
The transform 
\be
\hat f_{S_t}(s,u)=\lap{t}\lap{y} f_{S_t}(t,y)
\ee
is taken with respect to the two coordinates $t$ and $y$ with conjugate variables $s$ and $u$.
This yields
\bea
\fl
\hat f_{S_t}(s,u)
&=\sum_{n\ge0}\lap {t}\bigmean{\e^{-uS_t}I(t_n<t<t_n+\tau_{n+1})}
\nonumber\\
\fl
&=
\sum_{n\ge0}\bigmean{\e^{-u(\alpha_1\tau_1+\cdots+\alpha_n\tau_n)}\e^{u\alpha_{n+1}t_n}
\int_{t_n}^{t_n+\tau_{n+1}}{\rm d } t\,\e^{-st}\e^{-u\alpha_{n+1}t}},
\eea
where the average is taken upon the $\tau_i$ and $\alpha_i$ and $I(.)$ is the indicator random variable of the event inside the parentheses, equal to 1 if this event occurs and 0 otherwise.
The expression of the integral is
\be
\fl
\int_{t_n}^{t_n+\tau_{n+1}}{\rm d } t\,\e^{-t(s+u\alpha_{n+1})}=
\e^{-(s+u\alpha_{n+1})t_n}
\frac{1-\e^{-(s+u\alpha_{n+1})\tau_{n+1}}}{s+u\alpha_{n+1}},
\ee
thus
\beqa
\fl
\hat f_{S_t}(s,u)
&=\sum_{n\ge0}\bigmean{\e^{-u(\alpha_1\tau_1+\cdots+\alpha_n\tau_n)}\e^{-st_n}
\frac{1-\e^{-(s+u\alpha_{n+1})\tau_{n+1}}}{s+u\alpha_{n+1}}}
\\
\fl
&=\sum_{n\ge0}\bigmean{\hat\rho(s+u\alpha_1)\dots\hat\rho(s+u\alpha_n)\frac{1-\hat\rho(s+u\alpha_{n+1})}{s+u\alpha_{n+1}}}
\label{eq:fSsu}
\\
\fl
&=\sum_{n\ge0}\mean{\hat\rho(s+u\alpha)}^n\,
\bigmean{\frac{1-\hat\rho(s+u\alpha_{n+1})}{s+u\alpha}},
\eeqa
where, in the last two lines, the averages are taken upon the $\alpha_i$ only.

We thus finally obtain
\beq\label{eq:finaliid}
\hat f_{S_t}(s,u)
=\frac{1}{1-\mean{\hat\rho(s+u\alpha)}}\bigmean{\frac{1-\hat\rho(s+u\alpha)}{s+u\alpha}},
\eeq
where the averages are taken upon $\alpha$.
For a generic distribution $f_\alpha(a)$, we get
\beq\label{eq:continu}
\hat f_{S_t}(s,u)=
\frac{1}{1-\int{\rm d } a\, f_{\alpha}(a)\hat\rho(s+ua)}\int{\rm d } a\, f_{\alpha}(a)\,\frac{1-\hat\rho(s+ua)}{s+ua},
\eeq
while for the particular case of a discrete distribution (see (\ref{eq:faaD})), (\ref{eq:finaliid}) yields
\beq\label{eq:discret}
\hat f_{S_t}(s,u)=\frac{1}{1-\sum_{j=1}^q p_j\,\hat\rho(s+ua_j)}\sum_{j=1}^q p_j\frac{1-\hat\rho(s+ua_j)}{s+ua_j}.
\eeq

\subsection{Scaling regime}

In the long-time regime where $s$ and $u$ are small and comparable,
using (\ref{eq:spetit}), the result (\ref{eq:finaliid}) yields 
\beq\label{eq:scaling}
\hat{f}_{S_t}(s,u)\approx
\frac{\mean{(s+u\alpha)^{\theta-1}}}{\mean{(s+u\alpha)^{\theta}}}
\approx \frac{1}{s}g(\xi), \qquad \xi=\frac{u}{s},
\eeq
where
\be
g(\xi)= \frac{\mean{(1+\xi\alpha)^{\theta-1}}}{\mean{(1+\xi\alpha)^{\theta}}}.
\ee
If $\alpha$ is a continuous random variable we have
\beq\label{eq:gxiC}
g(\xi)=\frac{\int{\rm d } af_\alpha(a)(1+\xi a)^{\theta-1}}{\int{\rm d } af_\alpha(a)(1+\xi a)^{\theta}},
\eeq
while for the discrete case (\ref{eq:discret}) gives
\beq\label{eq:gxiD}
g(\xi)= \frac{\sum_{j}p_j(1+\xi a_j)^{\theta-1}}{\sum_{j}p_j(1+\xi a_j)^{\theta}}.
\eeq
The scaling behaviour (\ref{eq:scaling}) entails the following properties (see Appendix B of \cite{gl2001} for more details).
First, $S_t/t$ possesses a limiting distribution given by
\be
f_M(x)=\lim_{t\to\infty}f_{S_t/t}(t,x),\qquad x=\frac{y}{t}.
\ee
Hence
\be
\hat f_{S_t}(s,u)=
\int_0^\infty{\rm d } t\,\e^{-st}\mean{\e^{-uS_t}}
=\int_0^\infty{\rm d } t\,\e^{-st}\mean{\e^{-u tM}}
=\bigmean{\frac{1}{s+uM}},
\ee
so that
\beq\label{eq:gxidef}
g(\xi)=\left\langle\frac{1}{1+\xi M}\right\rangle=\int_{0}^{\infty}{\rm d } x\,\frac{f_{M}(x)}{1+\xi x}.
\eeq
This can be inverted as\footnote{
Setting $\xi=1/y$ in (\ref{eq:gxidef}) yields
\be
h(y)=\frac{1}{y}\,g\bigac{\frac{1}{y}}=\int_0^\infty{\rm d } x\,\frac{f_M(x)}{x+y},
\ee
showing that $h(y)$ is the Stieltjes transform of $f_M(x)$ \cite{widder}.
}
\be
f_{M}(x)=-\frac{1}{\pi x}\lim_{\epsilon\rightarrow 0}{\rm Im}\;
g\left(-\frac{1}{x+{\ii}\epsilon}\right).
\ee
Furthermore, the moments of $M$ can be obtained, when they exist, by expanding $g(\xi)$ as a
Taylor series, since (\ref{eq:gxidef}) implies that
\beq\label{eq:moments}
g(\xi)=\sum_{k\ge0}(-\xi)^{k}\left\langle M^{k}\right\rangle.
\eeq
We shall come back to the moments of $M$ in section \ref{sec:moments}. 

In the continuous case, the result is 
\beq\label{eq:fXx}
f_M(x)=\frac{\sin\pi\theta}{\pi}\frac{I_{\theta-1}^{<}(x)I_{\theta}^{>}(x)+I_{\theta}^{<}(x)I_{\theta-1}^{>}(x)}
{I_{\theta}^{<}(x)^2+I_{\theta}^{>}(x)^2+2\, I_{\theta}^{<}(x)I_{\theta}^{>}(x)\cos\pi\theta},
\eeq
with
\beq\label{eq:IthC}
I_{\theta}^{<}(x)=\int_{-\infty}^x{\rm d } a\,(x-a)^\theta f_{\alpha}(a),
\quad I_{\theta}^{>}(x)=\int_{x}^\infty{\rm d } a\,(a-x)^\theta f_{\alpha}(a).
\eeq
In the discrete case (see (\ref{eq:faaD})) we have in each sector $a_i<x<a_{i+1}$,
\beq\label{eq:IthD}
I_{\theta}^{<}(x)=\sum_{j\leq i} p_j(x-a_j)^{\theta},
\qquad I_{\theta}^{>}(x)=\sum_{j> i} p_j(a_j-x)^{\theta},
\eeq
which results in the expression for $f_M(x)$ given by (\ref{eq:central}), up to the replacement of $\pi_j$ by $p_j$.

Similar results can be found in \cite{barkai1,barkai2,barlow} (see section \ref{sec:discussion}).

\subsection{Examples}

Let us take, as a first example, the case where $f_\alpha$ is discrete (see (\ref{eq:faaD})), with $q=2$, and $p_1=p_2=1/2$.
Then, if $a_1<x<a_2$, (\ref{eq:fXx}) and (\ref{eq:IthD}) yield
\beqa\label{eq:lamperti}
\fl
f_M(x)=\frac{(a_2-a_1)\sin \pi\theta}{\pi}\frac{(x-a_1)^{\theta-1}(a_2-x)^{\theta-1}}
{(x-a_1)^{2\theta}+(a_2-x)^{2\theta}+2\cos\pi\theta(x-a_1)^\theta(a_2-x)^\theta},
\nonumber\\
\eeqa
and $f_M(x)=0$ otherwise, which is the law found by Lamperti \cite{lamperti}.
This function has a power-law singularity with negative exponent at both ends, $x\to a_j$ ($j=1,2$), 
\beq\label{eq:singular}
f_M(x)\approx\frac{\sin \pi\theta}{\pi(a_2-a_1)^\theta}|x-a_j|^{\theta-1}.
\eeq
It is U-shaped, as the arcsine law,
\be
f_M(x)=\frac{1}{\pi\sqrt{(a_2-x)(x-a_1)}},
\ee
to which it reduces when $\theta=1/2$, as long as $\theta<\theta_{\mathrm{c}}=0.594612...$, while a local maximum appears at $x=(a_1+a_2)/2$ when $\theta>\theta_{\mathrm{c}}$ \cite{balda,gl2001}.

As a second example, let the random variable $\alpha$ be uniform between $-1$ and $1$. 
Then, by (\ref{eq:IthC}), we have
\be
I_{\theta}^{<}(x)=I_{\theta}^{>}(-x)=\frac{(1+x)^\theta}{2(1+\theta)},
\ee
and therefore, if $-1<x<1$, (\ref{eq:fXx}) implies
\beq\label{eq:unif}
\fl
f_M(x)=\frac{2(1+\theta)\sin \pi\theta}{\pi\theta}
\frac{(1-x^2)^\theta}
{(1-x)^{2(1+\theta)}+(1+x)^{2(1+\theta)}+2\cos\pi\theta(1-x^2)^{(1+\theta)}},
\nonumber\\
\eeq
and $f_M(x)=0$ otherwise.
This function vanishes as a power-law at both ends, $x\to \pm1$, with a positive exponent 
\be
f_M(x)\approx\frac{(1+\theta)\sin \pi\theta}{2^{1+\theta}\pi\theta}(1\mp x)^{\theta},
\ee
and is always maximum at $x=0$.
Note that (\ref{eq:unif}) reduces to the arcsine law for $\theta=-1/2$.

Finally, it is easy to see on both expressions (\ref{eq:lamperti}) and (\ref{eq:unif}) that $f_M(x)\to f_\alpha(x)$ for $\theta\to0$ (complete absence of self-averaging), and that, when $\theta\to1$, $f_M(x)$ becomes a $\delta$ function centered at $\mean{\alpha}$, that is, at $(a_2+a_1)/2$ for the former and at 0 for the latter (ergodicity).
The same holds true for the general expressions (\ref{eq:central}) and (\ref{eq:fXx}), as can be seen on (\ref{eq:gxiC}) and (\ref{eq:gxiD}).
We shall come back to these limits and their interpretations in section \ref{sec:moments} (see also \cite{lamperti,dorn,newman,balda,gl2001,barkai1,barkai2}).

\section{Distribution of the sum $S_t$ for a Markov renewal process}
\label{sec:markov}

We now assume the Markov chain to be irreducible with the associated stationary probability measure 
$\bra{\pi}=(\pi_1,\ldots,\pi_q)$ satisfying
\beq \label{eq:measure}
\bra{\pi}=\bra{\pi}\m P,\quad\sum_{i=1}^q\pi_i=1.
\eeq

The main results of this section are, first, the exact expression (\ref{eq:key}) of the distribution $f_{S_t}$ of the sum $S_t$ in Laplace space, and secondly the scaling form (\ref{eq:gxi}) which leads to the limiting distribution (\ref{eq:central}) of the rescaled variable $M$ in real space.
This latter expression is the same as that founded for the iid case in section \ref{sec:iid}, up to the replacement of $\pi_j$ by $p_j$.
The difference between the Markov renewal process and the iid case is that the stationary distribution $(\pi_1,\ldots,\pi_q)$ is generated dynamically for the former, while the weights $p_i$ are given a priori for the latter.

\subsection{The distribution of the sum $S_t$}

We start again from (\ref{eq:fSsu})
\beq\label{eq:start}
\hat f_{S_t}(s,u)
=\sum_{n\ge0}\bigmean{\hat\rho(s+u\alpha_1)\dots\hat\rho(s+u\alpha_n)\frac{1-\hat\rho(s+u\alpha_{n+1})}{s+u\alpha_{n+1}}},
\eeq
where now the average is upon the configurations $\{\alpha_1,\alpha_2,\dots,\alpha_{n+1}\}$ of the chain.
A realisation of such a configuration, with $N_t=n$ fixed, is given by the sequence of values
\beq\label{eq:realisation}
a_{j_1},a_{j_2},\dots,a_{j_{n+1}},
\eeq
where each of the indices $j_1,j_2,\dots$ takes the values $1,\dots,q$.
Let $x_j$ and $y_j$ denote the quantities appearing in (\ref{eq:discret})
\be
x_j=\hat\rho(s+ua_j),
\qquad
y_j=\frac{1-\hat\rho(s+ua_j)}{s+ua_j}.
\ee
We also denote by $Q_{j_1}=\prob(\alpha_1=a_{j_1})$ the probability that the first value taken by $\alpha$ be
$a_{j_1}$.

Now (\ref{eq:start}) entails
\be
\hat f_{S_t}(s,u)=\sum_{n\ge0}\sum_{j_1,\dots,j_{n+1}}
Q_{j_1}\,x_{j_1}P_{j_1,j_2}\,x_{j_2}P_{j_2,j_3}\dots x_{j_n}P_{j_n,j_{n+1}}\,y_{j_{n+1}},
\ee
or, with matrix notations,
\be
\fl
\hat f_{S_t}(s,u)=\sum_{n\ge0}\sum_{j_1,\dots,j_{n+1}}
 Q_{j_1}\,\m X_{j_1,j_1}\m P_{j_1,j_2}\,\m X_{j_2,j_2}\m P_{j_2,j_3}\dots \m X_{j_n,j_n}\m P_{j_n,j_{n+1}}\,\m Y_{j_{n+1},j_{n+1}},
\ee
where we have introduced the diagonal matrices $\m X$ et $\m Y$,
\be
\m X=\mathrm{diag}(x_{1},\dots,x_q),\qquad \m Y=\mathrm{diag}(y_{1},\dots,y_q).
\ee
So
\beq\label{eq:final}
\fl
\hat f_{S_t}(s,u)
=\sum_{n\ge0}\sum_{j_1,j_{n+1}}
 Q_{j_1}(\m X \m P)^n_{j_1,j_{n+1}} \m Y_{j_{n+1},j_{n+1}}
=\sum_{n\ge0}\bra{Q}(\m{X}\m{P})^n\m Y\ket{R},
\eeq
with
\be
\ket{R}=\pmatrix {
1\cr
1\cr
\vdots \cr},
\qquad \bra{Q}=(Q_{1},\dots,Q_{q}),
\ee
so that $\braket{Q}{R}=1$.
Equation (\ref{eq:final}) finally leads to the key result
\beq\label{eq:key}
\hat f_{S_t}(s,u)= \bra{Q}(\m 1-\m{X}\m{P})^{-1}\m Y\ket{R}.
\eeq
For $u=0$, this expression yields
\be
\hat f_{S_t}(s,0)=\frac{1-\hat \rho(s)}{s} \bra{Q}(\m 1-\hat\rho(s)\m{P})^{-1}\ket{R}=\frac{1}{s},
\ee
showing that $f_{S_t}$ is well normalised.

\subsection{Scaling regime}
In the long-time regime where $s$ and $u$ are small and comparable, using again (\ref{eq:spetit}), we have
\be
\m X\approx \m 1-As^\theta \m D_\theta,
\ee
with
\be
\m D_\theta=\mathrm{diag}((1+\xi a_{1})^{\theta},\dots,(1+\xi a_{q})^{\theta}).
\ee
Likewise
\be
\m Y\approx As^{\theta-1}\m D_{\theta-1}.
\ee
The matrix $\m P$ is dominated by the Perron-Frobenius eigenvalue $1$, hence
the matrix $(1-\m{X}\m{P})^{-1}$ becomes singular when $s\to0$.
The final result reads
\beq\label{eq:result}
\hat f_{S_t}(s,u)=
 \bra{Q}(\m 1-\m{X}\m{P})^{-1}\m Y\ket{R}
\approx \frac{1}{s}\frac{\braopket{\pi}{\m D_{\theta-1}}{R}}{\braopket{\pi}{\m D_\theta}{R}}=\frac{1}{s}g(\xi),
\eeq
where
\beq\label{eq:gxi}
g(\xi)=\frac{\sum_{j}\pi_j(1+\xi a_j)^{\theta-1}}{\sum_{j}\pi_j(1+\xi a_j)^{\theta}},
\eeq
as we now show.

We write
\be
\m 1-\m X\m P\approx \m 1-(\m 1-As^\theta\m D_\theta)\m P\approx \m 1-\m P+As^\theta\m D_\theta\m P.
\ee
The matrix $\m T=\m 1-\m P$ has a zero eigenvalue, with associated (right and left) eigenvectors
\be
\ket{R},\qquad \bra{L}=(\pi_1,\pi_2,\dots)=\bra{\pi},
\ee
i.e.,
\be
\m T \ket{R}=0\qquad \bra{L}\m T=0.
\ee
For a generic matrix $\m G$, it is known that, $\epsilon$ being a small parameter,
\beq\label{eq:singulier}
(\m T+\epsilon\m G)^{-1}\approx\frac{1}{\epsilon}\frac{\ket{R}\bra{L}}{\braopket{L}{\m G}{R}}.
\eeq
Here, using (\ref{eq:singulier}), we get
\be
(\m T+As^\theta \m D_\theta\m P)^{-1}=\frac{1}{As^\theta}\frac{\ket{R}\bra{L}}{\braopket{L}{\m D_\theta\m P}{R}}
=\frac{1}{As^\theta}\frac{\ket{R}\bra{L}}{\braopket{L}{\m D_\theta}{R}},
\ee
since $\m P \ket{R}=\ket{R}$.
Thus
\be
\fl
\bra{Q}(\m1-\m{X}\m{P})^{-1}\m Y\ket{R}
\approx \braket{Q}{R} \frac{1}{s}\frac{\braopket{L}{\m D_{\theta-1}}{R}}{\braopket{L}{\m D_\theta}{R}}
= \frac{1}{s}\frac{\braopket{\pi}{\m D_{\theta-1}}{R}}{\braopket{\pi}{\m D_\theta}{R}},
\ee
which is (\ref{eq:result}).

Coming back to (\ref{eq:gxi}) we recognise the expression (\ref{eq:gxiD}) found previously,
up to the replacement of $p_j$ by $\pi_j$,
the stationary distribution.
As a consequence, the result for the distribution of the mean $M=\lim_{t\to\infty}S_t/t$
is the same as before (up to the replacement of $p_j$ by $\pi_j$), i.e., it is given by (\ref{eq:fXx}) and (\ref{eq:IthD}), resulting in (\ref{eq:central}).
The rationale behind this result is that the chain visits a great many times all accessible states.
Of course, as we shall see shortly, finite-time behaviours are different for the iid situation of section \ref{sec:iid} and for the Markov case of the present section.

\section{Moments}
\label{sec:moments}

\subsection{Moments in the long-time regime}
The moments of the mean $M$ can be obtained from (\ref{eq:gxi}), as mentioned above (see (\ref{eq:moments})).
For instance, the first three moments read
\beqa\label{eq:momentsM}
\mean{M}=\mean{\alpha},\qquad \mean{M^2}=\theta\mean{\alpha}^2+(1-\theta)\mean{\alpha^2},
\nonumber\\
\mean{M^3}=\theta^2\mean{\alpha}^3+\frac{3}{2}\theta(1-\theta)\mean{\alpha}\mean{\alpha^2}+
\frac{(1-\theta)(2-\theta)}{2}\mean{\alpha^3}.
\eeqa
These results manifest the absence of self-averaging of the process as long as $\theta<1$.
When $\theta\to0$, $M$ identifies to $\alpha$ (complete absence of self-averaging).
For $\theta=1$, the moments of $M$ are given by powers of $\mean{\alpha}$, namely $\mean{M^k}=\mean{\alpha}^k$.
More generally, if $\theta\ge1$, 
the system becomes ergodic in the limit of long times, i.e., the limiting distribution $f_M(x)$ is peaked around 
$\mean{\alpha}$, so
\be
M=\lim_{t\to\infty}\frac{1}{t}\int_{0}^t{\rm d } u\,\alpha(u)=\mean{\alpha},
\ee
i.e., the mean is identical to the average (see \cite{barkai1,barkai2} for similar considerations).
For $1\leq \theta<2$,
for long but finite times, even though
the distribution of $S_t/t$ becomes narrow, the fluctuations of $S_t$ are anomalous.
Finally, for $\theta\ge2$ they are normal and grow as $t^{1/2}$.
This phenomenon is analysed in detail in \cite{gl2001} for the case of two states. 
The present situation of a multistate Markov chain does not change this picture. 

\subsection{Finite-time corrections}

Coming back to the case where $\theta<1$, an interesting consequence of the analyses of sections \ref{sec:iid} and \ref{sec:markov} is the possibility of computing the finite-time corrections to the asymptotic formulas 
(\ref{eq:momentsM}), that is, in other words, of answering the question of how fast the fraction $M_t$ converges to its limit $M$, both for the iid case and for the Markov renewal process.
As we shall see, this convergence is quite slow, and different for the two processes.

We start from the exact expressions of $\hat f_{S_t}(s,u)$ given respectively by (\ref{eq:finaliid}) for the iid case and by (\ref{eq:key}) for the Markov renewal case. 

For the iid case, 
taking the derivative of (\ref{eq:finaliid}) with respect to $u$ and setting $u=0$, we have
\be
\lap{t}\mean{S_t}=\frac{\mean{\alpha}}{s^2},
\ee
yielding the identity, holding for any finite time $t$,
\beq\label{eq:identity}
\frac{\mean{S_t}}{t}=\mean{\alpha},
\eeq
which is in line with the result given in (\ref{eq:momentsM}) for $\mean{M}$.
This identity can also be simply obtained by noting that
\be
\mean{S_t}=\int_{0}^t{\rm d } u\, \mean{\alpha(u)}=\mean{\alpha}\big(\tau_1+\cdots+\tau_{N_t}+B_t\big)
=\mean{\alpha}t.
\ee

For the Markov renewal process, the identity (\ref{eq:identity}) no longer holds.
We have instead
\beq\label{eq:Mtasymp}
\frac{\mean{S_t}}{t}
\approx\mean{\alpha}+b\, t^{-\theta},
\eeq
where the amplitude $b$ of the correction is given by (\ref{eq:b}) below, as we now show.
We take the derivative of (\ref{eq:key}) with respect to $u$ and set $u=0$, to obtain, after some algebra,
\beq\label{eq:somme}
\lap{t}\mean{S_t}=\frac{1-\hat\rho(s)}{s^2}\bra{Q}(\m 1-\hat\rho(s)\m{P})^{-1}\m A\ket{R},
\eeq
where 
\be
\m A=\mathrm{diag}(a_{1},\dots,a_q).
\ee
Using the spectral decomposition of the matrix $\m P$, with eigenvalues $\la_i$ and right and left eigenvectors 
$\ket{R_i}$ and $\bra{L_i}$, 
\be
\m P=\sum_{i=1}^q\frac{\ket{R_i}\bra{L_i}}{\braket{L_i}{R_i}}\la_i,
\ee
we obtain
\be
(\m 1-\hat\rho(s)\m{P})^{-1}
=\sum_{i=1}^q\frac{1}{1-\hat\rho(s)\la_i}\frac{\ket{R_i}\bra{L_i}}{\braket{L_i}{R_i}}.
\ee
In the right side of this equation, the term coming from the Perron eigenvalue $\la_1=1$ plays a distinct role, so we rewrite it as
\be
(\m 1-\hat\rho(s)\m{P})^{-1}
=\frac{1}{1-\hat\rho(s)}\frac{\ket{R}\bra{L}}{\braket{L}{R}}+
\sum_{i=2}^q\frac{1}{1-\hat\rho(s)\la_i}\frac{\ket{R_i}\bra{L_i}}{\braket{L_i}{R_i}},
\ee
leading to the exact result, which is a more explicit expression of (\ref{eq:somme}),
\beq\label{eq:finalSt}
\lap{t}\mean{S_t}
=\frac{\mean{\alpha}}{s^2}+
\frac{1-\hat\rho(s)}{s^2}\sum_{i=2}^q\frac{1}{1-\hat\rho(s)\la_i}\frac{\braket{Q}{R_i}\braopket{L_i}{\m A}{R}}{\braket{L_i}{R_i}}.
\eeq
The first order correction is given by
\be
\lap{t}\mean{S_t}
\approx \frac{\mean{\alpha}}{s^2}+
As^{\theta-2}\sum_{i=2}^q\frac{1}{1-\la_i}
\frac{\braket{Q}{R_i}\braopket{L_i}{\m A}{R}}{\braket{L_i}{R_i}},
\ee
which, by inversion, yields (\ref{eq:Mtasymp}) with 
\beq\label{eq:b}
b=\frac{c}{\theta(1-\theta)}\sum_{i=2}^q\frac{1}{1-\la_i}\frac{\braket{Q}{R_i}\braopket{L_i}{\m A}{R}}{\braket{L_i}{R_i}},
\eeq
where $c$ is the tail coefficient of $\rho(\tau)$ (see (\ref{eq:power})).

These computations can in principle be extended to higher moments $\mean{(S_t/t)^k}$.
While they are easy for the iid case, they become increasingly more difficult for the Markov renewal process.
In any event, the finite-time corrections are again different for these two cases. 

We illustrate this study by the case of a symmetric simple random walk on $q=4$ sites, with reflecting boundary conditions.
The stationary probabilities of this Markov chain are $(\pi_1=1/6,\pi_2=1/3,\pi_3=1/3,\pi_4=1/6)$.
The random variable $\alpha$ of interest is the position of the walker, which takes the values $a_j=j$ ($j=1,\dots,4$).
With these values, the mean position of the walker is $\mean{\alpha}=5/2$ and the correction amplitude $b$ obtained from (\ref{eq:b}) reads 
\be
b=\frac{c}{\theta(1-\theta)}\Big\{-\frac{11}{4},-\frac{5}{4},\frac{5}{4},\frac{11}{4}\Big\},
\ee
according to whether the walker starts at $1,2,3,4$, respectively.

\begin{figure}[!ht]
\begin{center}
\includegraphics[angle=0,width=.7\linewidth,clip=true]{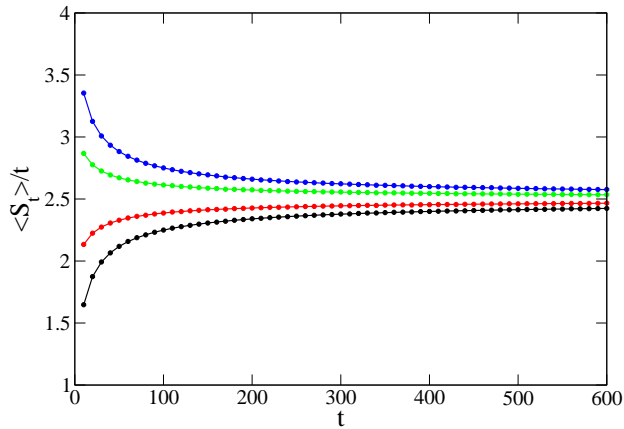}
\caption{\small
Plot of $\mean{S_t}/t$ against time $t$ for the four starting points $1,2,3,4$ of a simple random walker on $q=4$ sites, with reflecting boundary conditions.
Dots: simulation points, solid lines: numerical inversion of the exact Laplace transforms (\ref{eq:somme}) or (\ref{eq:finalSt}).
See text for details on $\rho(\tau)$.
Here $\theta=3/4$.
}
\label{fig:tout}
\end{center}
\end{figure}

\begin{figure}[!ht]
\begin{center}
\includegraphics[angle=0,width=.7\linewidth,clip=true]{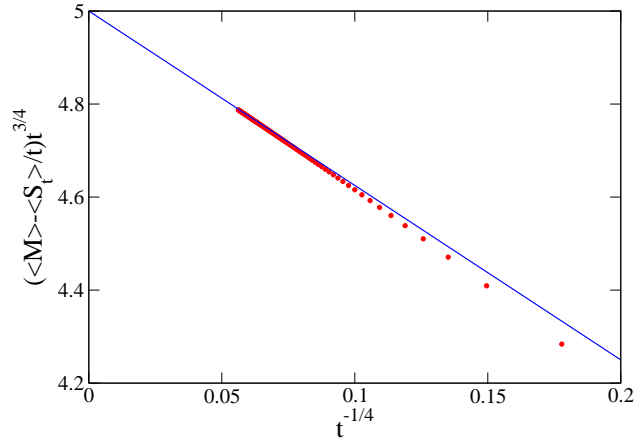}
\caption{\small
Check of the theoretical prediction of the corrections to scaling (\ref{eq:correction2}).
 Dots: left side of
 (\ref{eq:straight}) against $t^{-1/4}$,
obtained by a numerical inversion of the exact Laplace transforms (\ref{eq:somme}) or (\ref{eq:finalSt}).
 Solid line: straight line $y=5-15x/4$.
}
\label{fig:correction}
\end{center}
\end{figure}

Figures \ref{fig:tout} and \ref{fig:correction} depict a numerical study of this process.
The random time intervals $\tau$ are drawn from the distribution $\rho(\tau)=\theta/\tau^{1+\theta}$ for $\tau\ge1$, with tail coefficient $c=\theta$, corresponding to taking
$\tau=U^{-1/\theta}$, where $U$ is uniform between 0 and 1.
We choose $\theta=3/4$, yielding $b=\{-11,-5,5,11\}$, according to the initial position of the walker.
In figure \ref{fig:tout}, the agreement between the simulation points (dots) and the data coming from a numerical inversion of the exact expressions (\ref{eq:somme}) or (\ref{eq:finalSt}) of $f_{S_t}$ in Laplace space (solid lines) is excellent.
In figure \ref{fig:correction}, the convergence to the predicted amplitude $b=-5$, for a walker starting at $j=2$, is demonstrated by plotting the straight line $y=5-15x/4$, together with
\beq\label{eq:straight}
\bigac{\frac{5}{2}-\frac{\mean{S_t}}{t}}t^\theta\approx 5-\frac{15}{4}t^{\theta-1},
\eeq
against $t^{\theta-1}$.
Equation (\ref{eq:straight}) stems from the estimate
\beq\label{eq:correction2}
\frac{\mean{S_t}}{t}\approx \frac{5}{2}-\frac{5}{t^\theta}+\frac{15}{4t},
\eeq
obtained by expanding (\ref{eq:finalSt}) at second order.
The data were obtained by a numerical inversion of the exact expressions (\ref{eq:somme}) or (\ref{eq:finalSt}) of $f_{S_t}$ in Laplace space up to time $10^5$.
The agreement of these finite-time data with the theoretical prediction (\ref{eq:correction2}) is convincing.

Choosing a stable law for the distribution $\rho(\tau)$, with same tail parameter $c$ as above, would yield the same results, as can be seen on (\ref{eq:finalSt}) and (\ref{eq:b}).
In contrast, higher moments of $S_t$ depend on the details of the distribution $\rho(\tau)$.

\begin{figure}[!ht]
\begin{center}
\includegraphics[angle=0,width=.7\linewidth,clip=true]{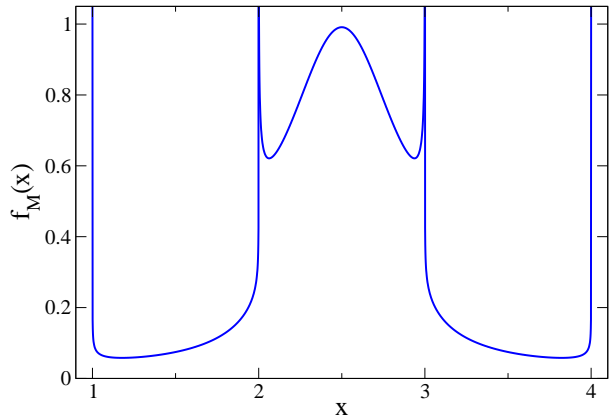}
\caption{\small
Stationary distribution $f_M(x)$ (see (\ref{eq:central})) for the simple random walk on $q=4$ sites with reflecting boundary conditions.
This function has power-law singularities with negative exponent $\theta-1$ at each integer $j=1,\dots,4$ ($\theta=3/4$).
}
\label{fig:fM}
\end{center}
\end{figure}

Finally, figure \ref{fig:fM} depicts the stationary distribution $f_M(x)$ (see (\ref{eq:central})) for this simple random walk.
This function has power-law singularities with negative exponent $\theta-1$ at each integer, i.e., when $x\to j$ ($j=1,\dots,4$), similarly to (\ref{eq:singular}).

%
\section{Discussion}
\label{sec:discussion}

The present work is part of the ongoing studies on the generalisations of the law of Lamperti for the occupation time of a two-state Markov process \cite{lamperti}.
Reviews in the mathematical literature of this topic can be found in \cite{pitman,james}.
The purpose of the present work was to extend this law to the specific case of a multistate semi-Markov process.
While completing this paper, we became aware of the existence of closely related works, with similar results \cite{barlow,pitman2,barkai1,barkai2}.
These studies are variations around the same theme, with differences, as we now comment.

Reference \cite{barlow} investigates the Walsh process of index $\theta$, defined as follows.
Consider $q$ half-lines $H_j$, $j=1,2,\dots,q$, with a common endpoint at zero, and a Bessel process (or radial Brownian motion) of dimension $2(1-\theta)$ on these half-lines.
When this process arrives at zero, it chooses the half-line $H_j$ with a given probability $p_j$.
Using the inherent scaling properties of Brownian motion, it is shown that the law of the rescaled sum $S_t/t$ obeys (\ref{eq:scaling}), (\ref{eq:gxiD}) and (\ref{eq:gxidef}).
The Walsh process therefore provides an implementation of the iid case of section \ref{sec:iid}, at least in the scaling regime (see also \cite{pitman2}).

References \cite{barkai1,barkai2} are closer in spirit to the present work.
The analysis of the process given in these references leads to the expression (\ref{eq:central}) of the distribution $f_{M}$ in the long-time regime, as well as to (\ref{eq:gxiD}), (\ref{eq:fXx}) and (\ref{eq:momentsM}).
Note that these results are already found in the iid case.
However the analysis made in \cite{barkai1,barkai2} does not lead to the explicit expressions of the distribution of the sum $S_t$ in Laplace space, as in (\ref{eq:finaliid}) for the iid case and in (\ref{eq:key}) for the Markov renewal case, which, in turn, lead to predictions of the finite-time behaviours of these processes, as demonstrated in section \ref{sec:moments}.

\ack It is a pleasure to thank J Pitman for useful correspondence and for pointing \cite{barlow} to us.

\end{document}